\documentclass{jpsj2}
\title{%
Two years of flight of the Pamela experiment: results and
perspectives. }

\author{%
Marco Casolino$^1$\thanks{E-mail address: casolino@roma2.infn.it},
Nicola De Simone $^1$, Daniel Bongue$^1$, Maria Pia De
Pascale$^1$,
 Valeria Di Felice$^1$,  Laura Marcelli$^1$, Mauro Minori$^1$,
Piergiorgio Picozza$^1$, Roberta Sparvoli$^1$, Guido
Castellini$^2$, Oscar Adriani$^3$, Lorenzo Bonechi$^3$, Massimo
Bongi$^3$, Sergio Bottai$^3$,  David Fedele$^3$, Paolo Papini$^3$,
Sergio Ricciarini$^3$, Piero Spillantini$^3$, Elena Taddei$^3$,
Elena Vannuccini$^3$,  Giancarlo Barbarino$^4$, Donatella
Campana$^4$, Rita Carbone$^4$,  Gianfranca De Rosa$^4$, Giuseppe
Osteria$^4$ Mirko Boezio$^5$, Valter Bonvicini$^5$, Emiliano
Mocchiutti$^5$, Andrea Vacchi$^5$,
 Gianluigi Zampa$^5$, Nicola Zampa$^5$,  Alessandro Bruno$^6$,
Francesco Saverio Cafagna$^6$, Marco Ricci$^7$, Petter
Hofverberg$^8$, Mark Pearce$^8$, Per Carlson$^8$, Edward
Bogomolov$^9$, S.Yu.~Krutkov$^9$, N.N.~Nikonov$^9$,
G.I.Vasilyev$^9$,  Wolfgang Menn$^{10}$, Manfred Simon$^{10}$,
 Arkady M. Galper$^{11}$, Lubov Grishantseva$^{11}$,  Sergey Koldashov$^{11}$, Alexey
 Leonov$^{11}$,
   Vladimir V.~Mikhailov$^{11}$,   Sergey A.~Voronov$^{11}$,  Yuri T.~Yurkin$^{11}$, Valeri
   G.~Zverev$^{11}$, Galina A. Bazilevskaya$^{12}$, Alexander N. Kvashnin$^{12}$,  Osman Maksumov$^{12}$,
   Yuri Stozhkov$^{12}$
}
\inst{%
 {$^1$  INFN and
Physics Department of University of Rome ``Tor Vergata" }, {$^2$
\em IFAC, Florence, Italy },
  {$^3$ \em INFN  and Physics Department of University of Florence
  },
  {$^4$\em INFN  and Physics Department of University of Naples ``Federico II"
  },
{$^5$\em INFN,  and Physics Department of University of Trieste},
{$^6$ \em INFN,  and Physics Department of University of Bari},
{$^7$ \em INFN, Laboratori Nazionali di Frascati,  Italy},
{$^8$\em KTH,   Stockholm, Sweden},
 $^9${\em Ioffe Physical Technical Institute, St. Petersburg,
 Russia},
{$^{10}$ \em Universit\"{a}t Siegen,    Germany },
 {$^{11}$\em Moscow Engineering and Physics Institute,
 Moscow, Russia},
{$^{12}$ \em Lebedev Physical Institute, Moscow, Russia }

} \recdate{\today}

\abst{%
\pam\ is a satellite borne experiment designed to study with great
accuracy cosmic rays of galactic, solar, and trapped nature in a
wide energy range (protons: 80 MeV-700 GeV, electrons 50 MeV-400
GeV). Main objective is the study of the antimatter component:
antiprotons (80 MeV-190 GeV), positrons (50 MeV-270 GeV) and
search for antinuclei with a precision of the order of $10^{-8}$).
The experiment, housed on board the  Russian Resurs-DK1 satellite,
was launched on June,
  $15^{th}$ 2006 in a $350\times 600~km$ orbit with an inclination of
70 degrees.    In this work we describe the   scientific
objectives and the performance of \pam\  in its first two years of
 operation. Data  on protons of trapped, secondary and galactic
 nature - as well as measurements of the December  $13^{th}$ 2006 Solar Particle Event - are also
 provided.
\\
{\it  To appear  on J. Phys. Soc. Jpn. as part of the proceedings
of the International Workshop on Advances in Cosmic Ray Science
March, 17-19, 2008 Waseda University, Shinjuku, Tokyo, Japan.}

 }

\kword{%
cosmic rays, antimatter, dark matter, solar particle events,
trapped cosmic rays }

\newcommand{\pam}{\textsf{PAMELA} }
\newcommand{\antip}{$\overline{p}$}
\newcommand{\posit}{$e^+$}

\newcommand{\grl}{Geophysical Research Letters}
\newcommand{\prd}{Physical Review D}
\newcommand{\apj}{ApJ}

\begin{document}
\maketitle

\section{Introduction}

\label{sec:intro} The scientific program of the Wizard
collaboration is devoted to the study of cosmic rays through
balloon and satellite-borne devices. Aims of this research
involve the precise determination of the antiproton \cite{boe97}
and positron \cite{boe00} spectrum, search of antimatter,
measurement of low energy trapped and solar cosmic rays with the
NINA-1 \cite{nina} and NINA-2 \cite{nina2} satellite experiments.
Other research on board Mir and International Space Station has
involved the measurement of the radiation environment,  the
nuclear abundances and the investigation of the Light Flash
\cite{nat} phenomenon with the Sileye experiments \cite{sil2,
sil3}. \pam\ is the largest and most complex device built insofar
by the collaboration, with the broadest scientific goals.  In this
work we describe the detector and its performance in the first two
years of operations.  Scientific objectives are presented together
with the   report of the first observations of protons of solar,
trapped and galactic nature.

\section{Instrument Description}
\begin{figure}
\begin{center}
\includegraphics[width=1.\textwidth]{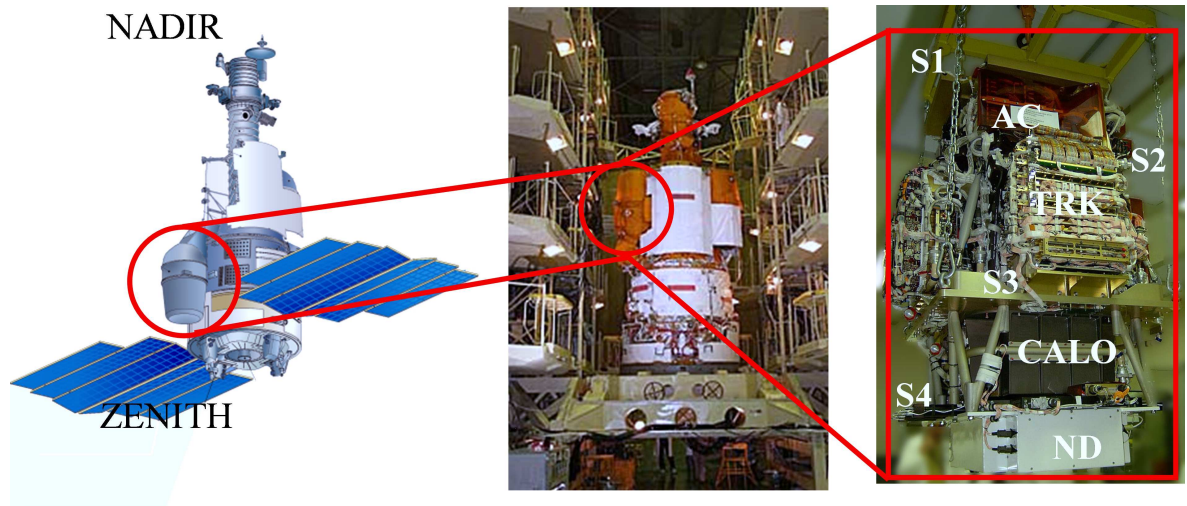}
\caption{Left: Scheme of the Resurs-DK1 satellite.  \pam\  is
located in the pressurized container on the left  of the picture.
In the scheme the pressurized container is in the acquisition
configuration. Center: The Resurs-DK1 satellite during integration
in Samara. The pressurized container housing Pamela is in the
folded (launch) position. Right: Photo of the \pam\ detector in
Tor Vergata with marked the position of the detectors. S1, S2, S3,
S4: scintillator planes; AC: top anticoincidence; TRK: tracker
core; CALO: Silicon-Tungsten calorimeter; ND: Neutron Detector.}
\label{schema}
\end{center}
  \end{figure}

 The device
(Figure \ref{schema}) is constituted by a number of highly
redundant detectors capable of identifying particles providing
charge, rigidity and velocity information over a very wide energy
range. A more detailed description of the device and the data
handling  can be found in { }\cite{Pi07, cpu, yoda}{ }. The
instrument is built around a  permanent magnet with a silicon
microstrip tracker and a scintillator system to provide trigger,
charge and time of flight information. A silicon-tungsten
calorimeter is used to perform hadron/lepton separation. A shower
tail catcher and a neutron detector at the bottom of the apparatus
are also employed to improve this separation. An anticounter
system  off line rejects spurious events produced in the side of
the main body of the satellite.
 Around the detectors are
housed the readout electronics, the interfaces with the CPU and
all primary and secondary power supplies. All systems (power
supply, readout boards etc.)  are redundant with the exception of
the CPU which is more tolerant to failures. The system is enclosed
in a pressurized container (Figure \ref{schema}) located on one
side of the Resurs-DK1 satellite. In a twin pressurized  container
is housed the   Arina experiment, devoted to the study of the low
energy trapped electron and proton component. Total weight of
\pam\ is 470 kg; power consumption is 355 W, geometrical factor is
21.6$\,cm^2 sr$.

\section{Antimatter component in cosmic rays and search for Dark Matter}
The study of the antiparticle component (\antip, \posit)
  of cosmic rays  is the main scientific goal of \pam.  A long term and detailed study of the antiparticle spectrum over a very wide energy
  range
 will allow to shed light over several questions of cosmic ray physics,  from particle production and propagation in the galaxy to
 charge dependent modulation in the heliosphere to dark matter
 detection. See \cite{pamcasopico08,pamcaso08} for a discussion of the antimatter
 detection capabilities of \pam.

\section{Measurement of cosmic rays in Earth's magnetosphere}

\begin{figure}[ht]
\begin{center}
\includegraphics[width=1.\textwidth]{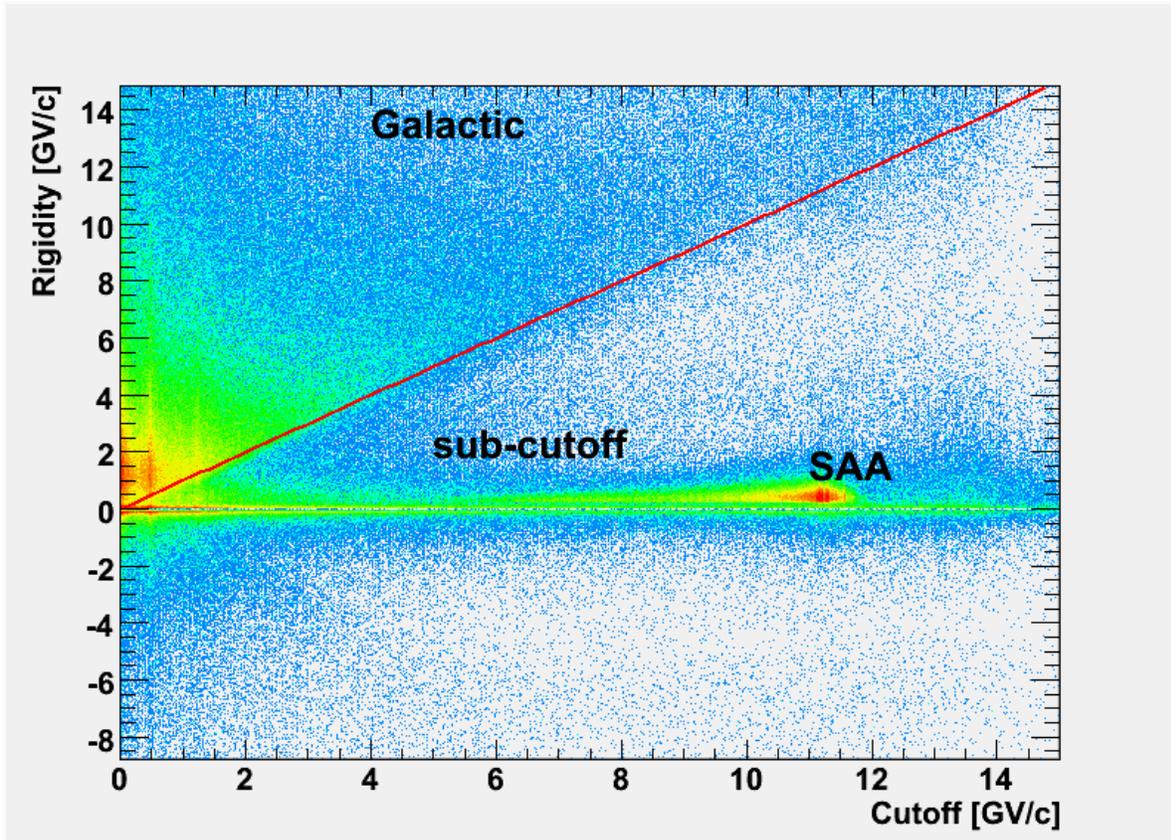}
\end{center}
\caption{Histogram of the rigidity $R_{tr}$ measured in the
tracker vs vertical Stormer Cutoff. Particles with positive charge
(p, $e^+$) have $R_{tr}>0$ and particles with negative charge have
$R_{tr}<0$. The effect of the geomagnetic field on galactic
particles is clearly visible. Primary particles, of galactic or
solar origin, have a rigidity  above the local Stormer cutoff (see
text) and are well separated from reentrant albedo events (below
the cutoff) produced in the interaction of primaries with the
Earth's atmosphere. It is also possible to see the spot of high
fluence of low ($R<2$ GV) protons trapped in the inner Van Allen
belt, crossed by \pam\ in the South Atlantic Anomaly (SAA)
region.}\label{rigcutoff}
\end{figure}

Earth's magnetic field can be used as a spectrometer to separate
cosmic rays of various nature and origin.  To   separate the
primary (galactic) component from the reentrant albedo (particles
produced in interactions of cosmic rays with the atmosphere below
the cutoff and propagating along Earth's magnetic field line)
component it is necessary to evaluate the local geomagnetic
cutoff. This is estimated using the IGRF magnetic field model
along the orbit; from this the McIlwain $L$ shell is
calculated\cite{igrf}. In this work we have used the vertical
Stormer (defined as $G=14.9/L^2$) approximation\cite{shea} to
separate between particles of different nature. Figure
\ref{rigcutoff} shows the rigidity of particles as function of the
evaluated cutoff $G$. The primary (galactic) component, with
rigidities above the cutoff is clearly separated from the
reentrant albedo (below cutoff) component, containing also trapped
protons in the South Atlantic Anomaly (SAA).

\subsection{Solar modulation of GCR}

Launch of \pam\ occurred during the XXIII solar minimum. At solar
minimum the magnetic field of Sun has an approximatively dipolar
structure, currently with negative polarity (A$<$0, with magnetic
field lines directed toward the sun in the northern emisphere). We
are currently in an  unusually long solar minimum with various
predictions on the behavior of the intensity and peaking time of
next maximum.  In the 2006-2008  period \pam\ has been observing
an increase of the flux of  galactic cosmic rays at low energy
($<1 $  GeV) due to solar modulation caused by the decreasing
solar activity. A long term measurement of the behaviour of the
proton, electron and $Z\geq 2 $  flux at 1 AU can   provide
information on propagation phenomena occurring in the heliosphere.
The possibility to measure the antiparticle spectra will allow to
study also charge dependent solar modulation effects.

\begin{figure}[ht]
\begin{center}
\includegraphics[width=1.\textwidth]{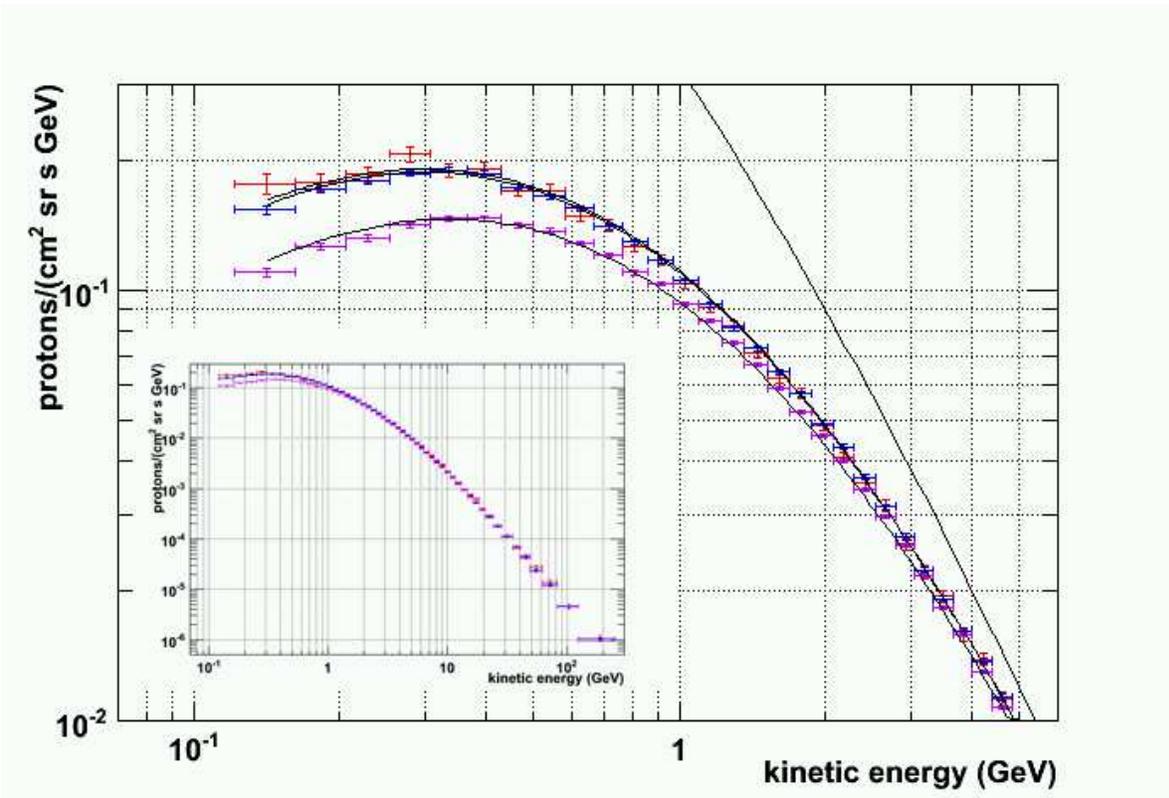}
\end{center}
 \caption{Differential spectrum of protons measured in July 2006 (purple - bottom),
  August 2007 (black - central), February 2008 (red - top curve). Below 1 GeV it is possible to see the  effect of solar modulation on the
         flux variation.
The straight black line represent the assumed interstellar
spectrum. Only statistical errors are shown. }
\label{solmodulation} \vspace{-.5cm}
\end{figure}

The MDR (Maximum Detectable Rigidity) of the magnet spectrometer
is  $ \sim 1 GTV$ ($700$ GV on average) allows to measure the
spectrum of cosmic-ray protons from 80~MeV up to almost 1~TeV; in
this work we present proton data up to 200 GeV. Proton fluxes have
been obtained requiring a clean track hitting the scintillator and
fitted in the tracker  with  energy loss compatible with protons
(rejecting He nuclei and secondary pions produced in the
satellite). Particles of galactic origin are selected requiring
that the rigidity of the event $R$  is above the local cutoff
($R>G*1.3 $) to avoid contamination of the secondary component. To
evaluate absolute spectra it was necessary to take into account
live time, geometrical factor and detector efficiencies, using
Montecarlo simulations (Geant 3.21) to evaluate the efficiency of
each cut at various energies for each detector configuration. A
compared  study of the temporal and energetic variations of the
efficiencies with experimental data is currently in progress. The
current approach with Montecarlo simulations has an associated
systematic error estimated of the order of $10\%$ not shown in
figures and tables. In Figure \ref{solmodulation} are shown the
proton fluxes measured in various periods of the mission. The
 effect of decreasing solar activity on the increasing flux of cosmic rays
is visible even at solar quiet period,   in agreement with the
increase of neutron monitor fluxes. From the  flux  $J(E,t)$ it is
possible to evaluate the solar modulation parameter $\Phi (t) $.
The heliosphere is thus approximated with a spherical
structure\cite{solmodulationgleeson1968},
  assuming
that particles lose energy independently from the sign of the
charge and incoming direction  to enter the heliosphere according
to the following:
\begin{equation}
J(E,t)= \frac{E^2 - E_0^2}{(E+|Z|e\Phi(t))^2-E_0^2}
J_{is}(E+|Z|e\Phi(t))
\end{equation}

More detailed models   involve correlation of the particle flux
and solar modulation with variation with time of tilt angle of the
heliospheric current sheet.    In this work we have assumed a
dependence of the interstellar spectrum according to
\cite{bess2007}:
\begin{equation}
J_{is} = A   \beta^{0.7}_{is} R_{is}^{-\gamma}
\end{equation}
With $\beta=v/c$ and $v$ the speed of the particle. The value of
$\gamma $ is obtained from the fit at high energies (from 15-20 to
200 GeV), where solar modulation effects become negligible. For
\pam\ we obtain $\gamma = 2.76 \pm 0.01$. The estimation of the
value of $A $ with a precision required to estimate $\Phi $ is
more complex and can affect the determination of the absolute
value of $\Phi $. In table \ref{solmod} are shown the values of
$\Phi (t) $ obtained with different assumptions  of the value $A$.
 All values used are compatible with the various fits and - even
though differ  by as little as $2.5 \% $  - produce different
values of $\Phi (t) $. It should be noted, however, that the
decrease of the modulation parameter  from 2006 to 2008
$\Delta\Phi_{ij}=\Phi(t_i)- \Phi(t_j) $  depends less from the
assumption of $A $ and can be considered more reliable. The values
of $A$ are shown here only to show the effect on the absolute
value of solar modulation and should not be used for the
evaluation of the  interstellar spectrum. A more detailed
estimation of $A$, using the full dataset of \pam is currently in
progress.

\subsection{Trapped particles in the Van Allen Belts }

\begin{figure}[ht]
\begin{center}
\includegraphics[width=.7\textwidth]{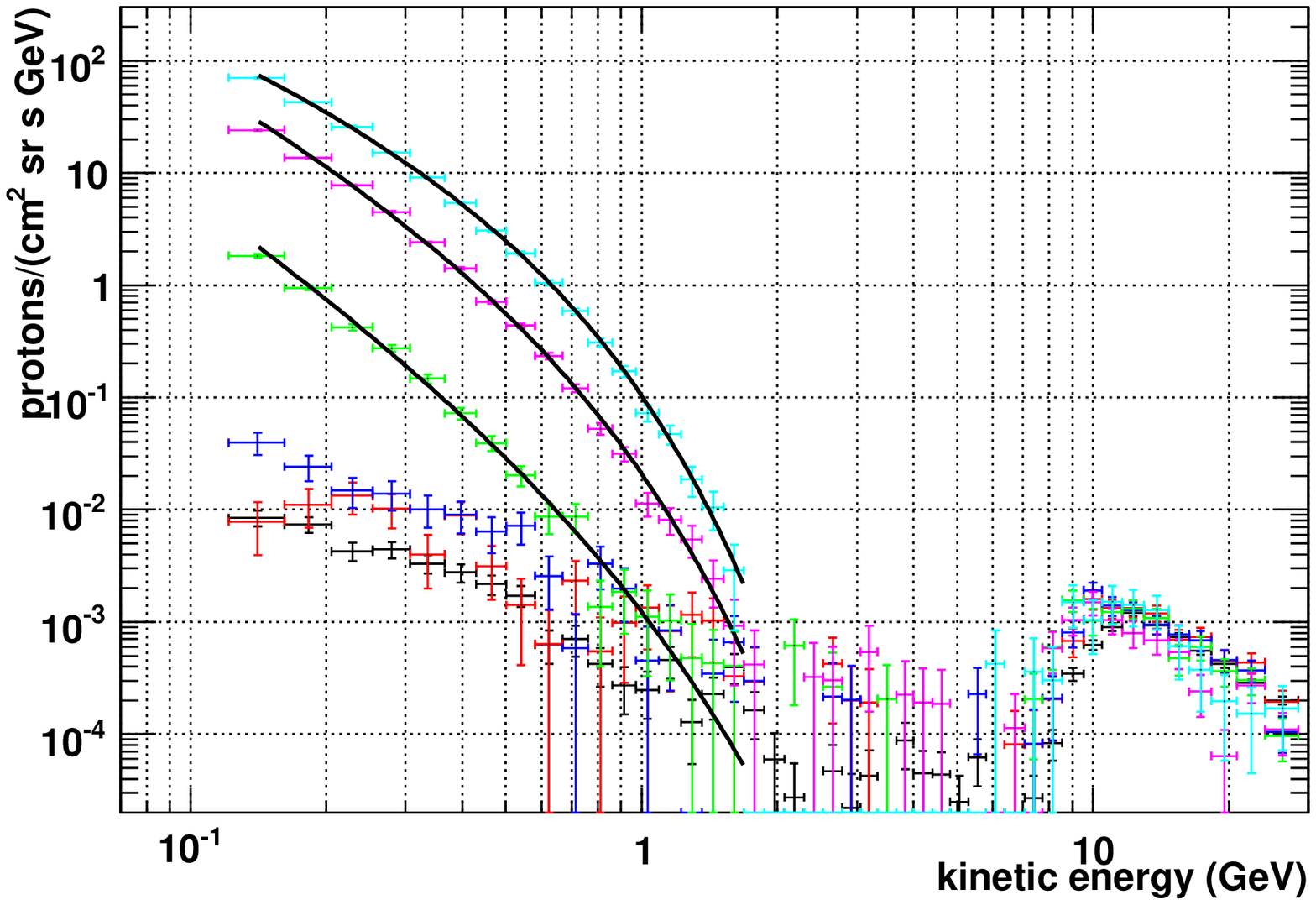}

\includegraphics[width=.7\textwidth]{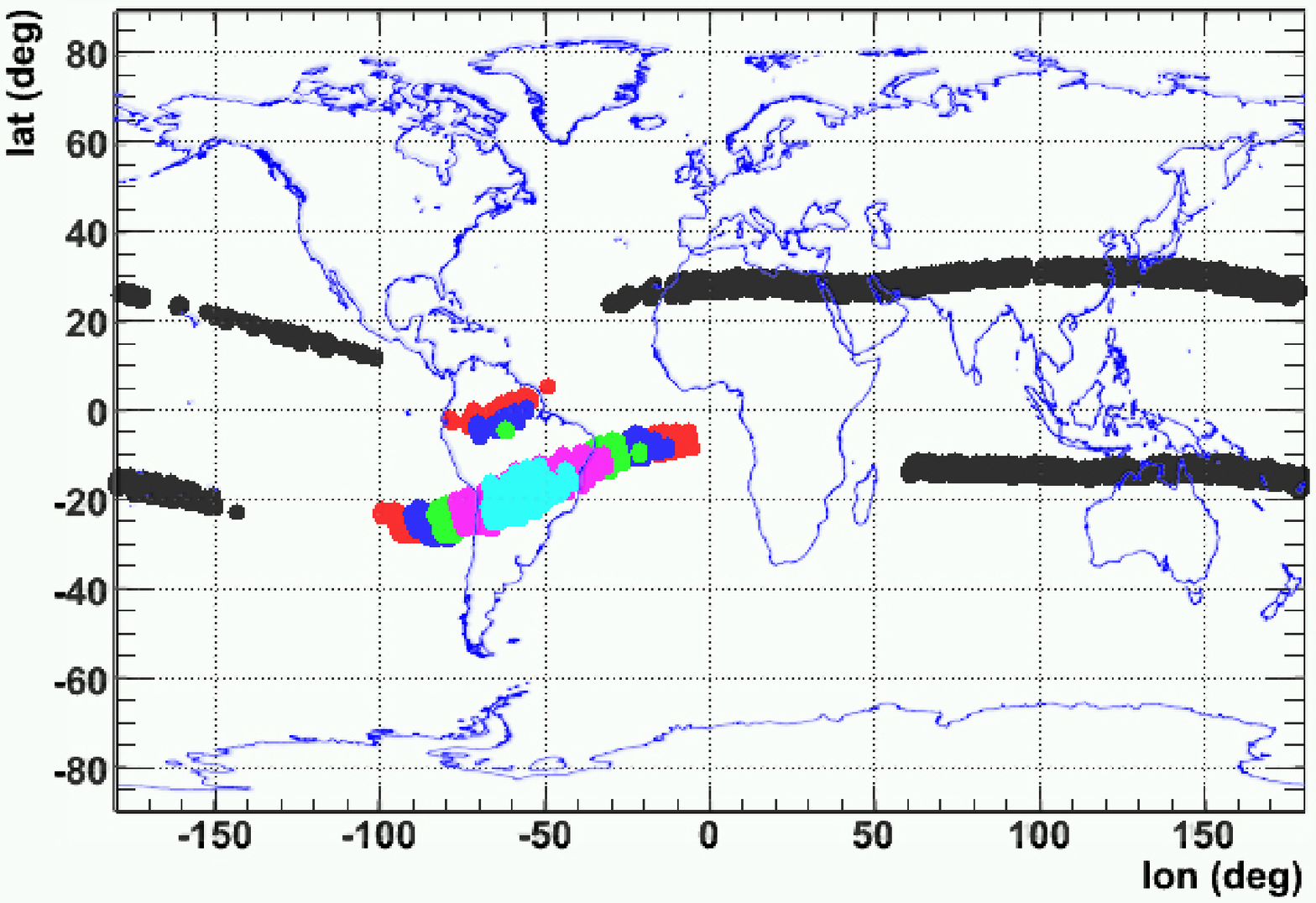}
\end{center} \caption{ Top: Plot of the differential energy spectrum of
\pam\ in different regions of the South Atlantic Anomaly.
Selection regions (shown in top bottom  panel) are selected
according to decreasing intensity of the magnetic field  from
bottom to top: Black $B
> 0.3 G$ - outside the SAA,
Red      $  0.22 $ G  $ < B < 0.23  $ G,
 Blue      $ 0.21$ G  $ < B < 0.22  $ G,
 Green       $ 0.20 G $ < B < 0.21  $ G,
 Pink     $  0.19  G  $< B < 0.20   $ G,
 Turquoise    $0.19 $ G > B )  in the  cutoff region
$10.8$  GV $< G < 11.5$ GV. Flux of trapped  particles can exceed
the
 secondary
particle flux in the same cutoff region outside the anomaly (black
bands) of about four orders of magnitude at low energy.
  }
 \label{southatlantic}
\end{figure}

The high energy ($>80 MeV$) component of the proton belt, crossed
in the South Pacific region can   be monitored in detail with \pam
.   In Figure \ref{southatlantic}  is shown the differential
energy spectrum measured in different regions of the South
Atlantic Anomaly. Proton selection criteria are the same used in
the determination of the absolute galactic spectrum.  It is
possible to see the increase of the flux toward the centre of the
anomaly. Particle flux exceeds several orders of magnitude the
flux of secondary (reentrant albedo) particles measured in the
same cutoff region outside the anomaly and it is maximum where the
magnetic field is lowest. The trapped component at the center  can
be fitted with energy dependent power law spectrum of the form
$\phi = A E^{-\gamma - \delta E }$. This measurement can be used
to validate various existing models\cite{ae8, mewaldtprotons}
providing information on the trapping and interaction processes in
Earth's magnetosphere. These studies will be expanded to address
temporal and spatial variations as well as different particle
species such as antiprotons\cite{mewaldtantiprotons}. These
results can be scaled to larger
 but less directly accessible -  magnetospheres such as
Jupiter or pulsars. In Table \ref{fitsaa} are shown the values
obtained  fitting the trapped spectra with a rigidity dependent
power law $\phi_{tr} =  A R^{-\gamma - \delta R} $.
%In the anomaly the dead  time of the instrument is higher due to the higher %particle flux but

\subsection{Secondary particles production in the Earth's
atmosphere}

\begin{figure}[ht]
\begin{center}
\includegraphics[width=1.\textwidth]{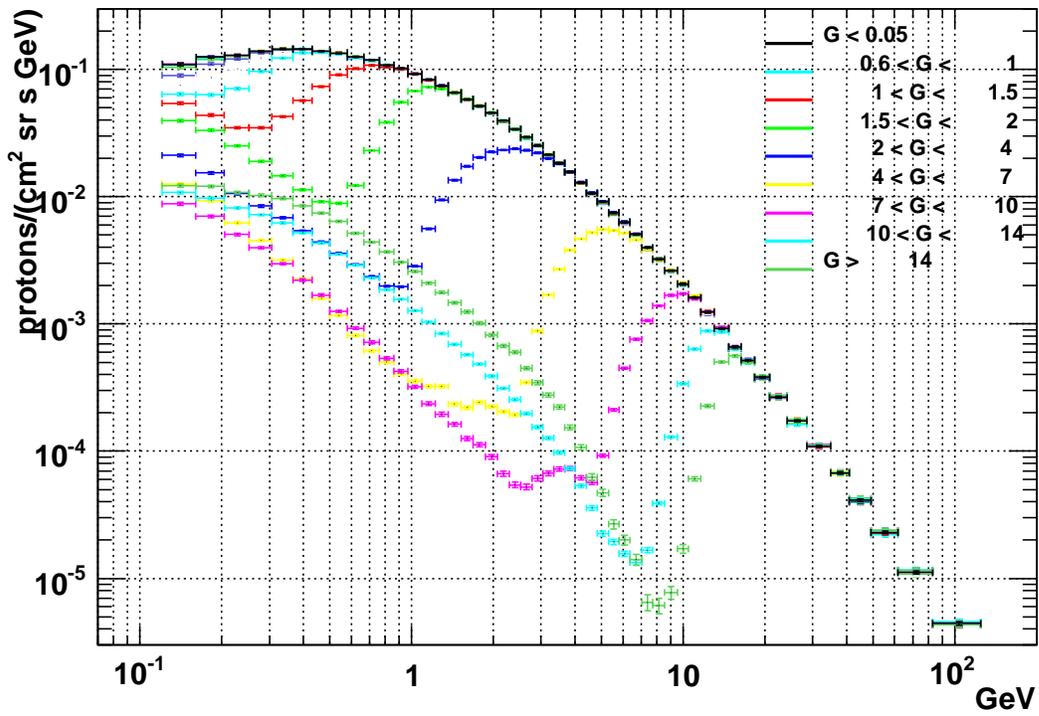}
\end{center}
\caption{Plot of the differential energy spectrum of \pam\ at
different values of geomagnetic cutoff $G$. It is possible to see
the primary spectrum at high rigidities and the reentrant albedo
(secondary) flux at low rigidities. Only statistical errors are
shown.} \label{subcutoff}
\end{figure}

 In Figure \ref{subcutoff}
is shown the particle flux measured in different cutoff regions.
It is possible to see  the primary (galactic - above cutoff)  and
the secondary (reentrant albedo - below cutoff ) component. At the
poles, where field lines are open and cutoff is below the minimum
detection threshold of \pam\ the secondary component is not
present. Moving toward lower latitude regions the cutoff increases
and it is possible to see the two components, with the position of
the gap increasing with the increase of the cutoff. The secondary
component of cosmic rays contributes to the atmospheric neutrino
production\cite{honda2004}. Therefore an accurate measurement of
the secondary component is of relevance in the reduction of the
uncertainties of the expected flux on the ground\cite{honda2007}
and   in the estimation of hadronic cross sections (protons on O
or N) at high energies, not otherwise determinable on ground.

\section{Solar energetic particles}

\pam\ observations are currently taking place at the solar minimum
of  XXIII cycle. Currently only one series of major solar events
 in December 2006  has been detected. It is
expected that in the next years, going toward XXIV cycle solar
maximum, more events will be detected by the apparatus. The
characteristics of \pam\ allow   real time measurements of
different particle spectra, important in understanding the
acceleration and propagation mechanisms which take place at the
Sun and in the heliosphere. In this section we briefly discuss the
observation capabilities of \pam\ and report on the proton spectra
observed in the December event.

\subsection{Electrons and positrons}
\label{sec:positron} Positrons are produced mainly in the decay of
$\pi^{+}$ coming from nuclear reactions occurring at the flare
site. Up to now, they have only been measured indirectly by remote
sensing of the gamma ray annihilation line at 511~keV. Using the
magnetic spectrometer of \pam\ it will be possible to separately
analyze the high energy tail of the electron and positron spectra
at 1 Astronomical Unit (AU) obtaining information both on particle
production and charge dependent propagation in the heliosphere in
perturbed conditions of  Solar Particle Events.

\subsection{Protons and nuclei}
\label{sec:proton} \pam\  can  measure the solar component over  a
very wide energy range (where the upper limit will be determined
by the size and  the spectral shape of the event). These
measurements can be correlated with other instruments placed in
different points of the Earth's magnetosphere to give information
on the acceleration and propagation mechanisms of SEP (Solar
Energetic Particle) events. Up to \pam measurements there has been
no direct measurement~\cite{miroshnichenko} of the high energy
($>$1~GeV) proton component of SEPs. Until now the spectrum of
solar energetic particles covering the energy range from $\simeq $
MeV/n to at least several GeV/n could not be measured by a single
device; spacecraft, balloon, and neutron monitor data currently
being used for this purpose \cite{galla}. Moreover, the energy
range of onboard spectrometers have some problems with $ >$ 100
MeV energy channels due to particles penetrating the detector from
the  side\cite{ssss}. Since the  majority of solar energetic
particle events have a spectrum turnover around ~100
MeV/n~\cite{ryan} it is extremely important to measure the whole
energy range of solar energetic particles by a single instrument.
Another important task is to find the upper limit of acceleration
processes at the Sun.
%The importance of a direct measurement of this spectrum is related
%to the fact~\cite{ryan} that there are many solar events where the
%energy of protons is above the highest ($\simeq$100 MeV)
%detectable energy range of current spacecrafts,  but is below the
%detection threshold of ground Neutron
%Monitors~\cite{bazilevskaya}. With \pam\   it will be possible to
%examine the turnover of the spectrum, where we find the limit of
%acceleration processes at the Sun.
Also the light nuclear component related to SEP events over a wide
energy range can be investigated. This should contribute to
establish whether there are differences between the compositions
of the high energy (1 GeV)
 and  the low energy component ($\simeq$ 20 MeV) ions   producing $\gamma $ rays or the quiescent solar corona\cite{ryan05}.
 These
measurements will help us to better understand the selective
acceleration processes in the higher energy impulsive
events~\cite{reames}.

\subsection{13 December 2006 Solar Particle Event}

At the time of writing the most significant events detected by
\pam\ occurred between December 6$^{th}$ and 17$^{th}$  2006 and
were originated from  active  region NOAA 10930. Dec 6$^{th}$
event was originated in the  East limb, resulting in a gradual
proton event reaching Earth on Dec 7$^{th}$ and lasting until the
events of Dec 13 and 14\cite{goes}.  On   13$^{th}$ December 2006,
02:38 UT  an X3.4/4B solar flare occurred   in active region NOAA
10930 ($S06^oW23^o$).
   The interaction   between the fast
rotating sunspot and the ephemeral regions triggers continual
brightening and finally produces the major
flare\cite{zhangsongsep2006}. The intensity of the event  is quite
unusual for a solar minimum condition. Starting at 2:50 UT on the
same day  various neutron monitors, with cutoff rigidities below
about $4.5\,GV$, recorded a Ground Level Enhancement (GLE70) with
relative increases ranging from $20\%$ up to more than $80\%$
(Apaty, Oulu) \cite{Bi07, Ta07}.  Apaty and Oulu also registered
the peak of the event beetween 02:40 UT and 03:10 UT, while most
of the neutron monitors had it between 03:10 UT and 03:40 UT.
The spectrum and its dynamic was investigated at higher energies
using ground measurements by neutron monitors at different cutoff
rigidities \cite{Va07} resulting in a spectral estimation assuming
a power law in rigidity of $\gamma \simeq 6$. The onset time was
later for the proton channels on-board of GOES-11 satellite: 03:00
UT for greater than 100 MeV protons and 03:10 for greater than 10
MeV protons \cite{Ta07}. \pam was in an high cutoff region at the
flare occurrence and reached the South Polar region   at about
03:10 UT. Muon monitors were also able to detect the GLE event and
its spatial-angular anisotropy has been measured \cite{Ti07}.
Differential proton spectra were directly meausured by GOES, ACE,
Stereo, SAMPEX at energies below $400\, MeV$.  With these
instruments it was also possibile to measure the elemental
composition of the various events\cite{Me07, Co07}.
\par  The
event produced also a full-halo Coronal Mass Ejection (CME) with a
projected speed in the sky of 1774 km/s \cite{mmcme}. The forward
shock of the CME reached Earth at 14:38 UT on December 14, causing
a Forbush decrease of galactic cosmic rays which lasted for
several days.

\begin{figure}[ht]
\begin{center}
\includegraphics[width=1.\textwidth]{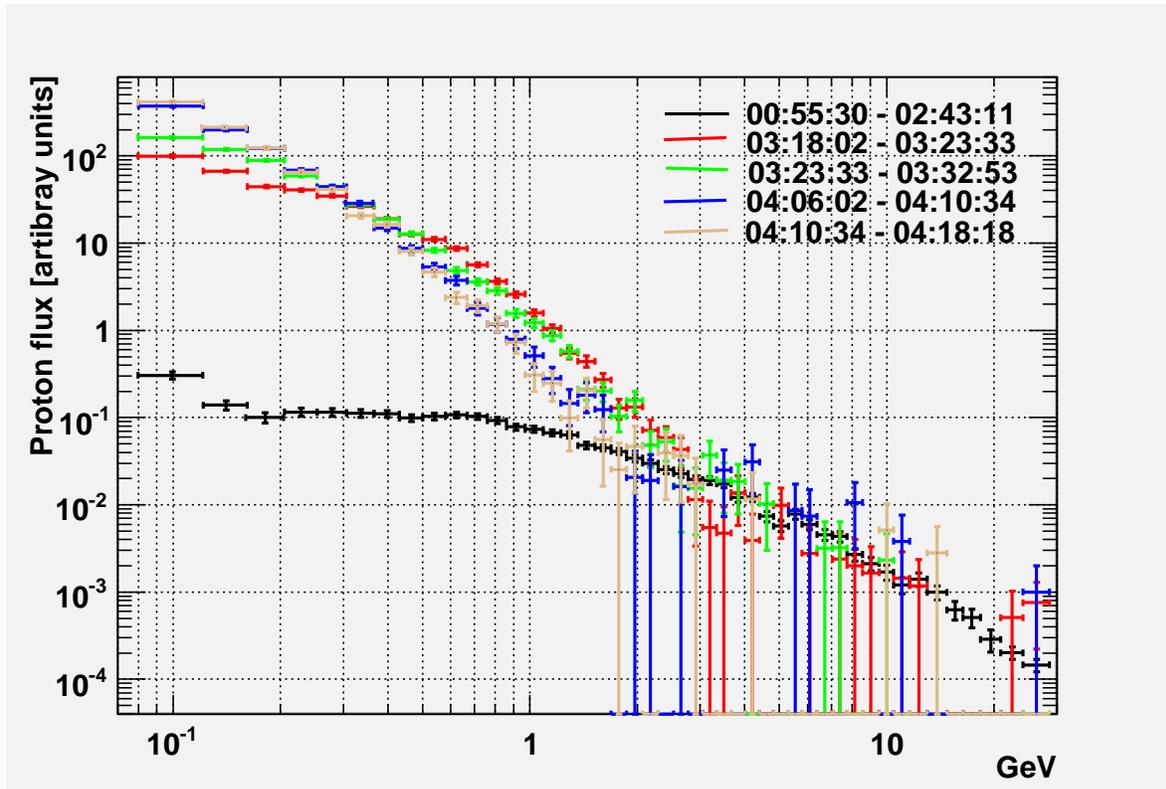}
\end{center}
\caption{Proton differential energy spectra (flux vs kinetic
energy) in different time intervals during the event of the 13th
December 2006.  The black line is the spectrum before the arrival
of the charged particles with a small peak at low energy due to
the presence of solar protons from previous events.  It can be
observed that the maximum flux of the high energy component of the
solar protons arrives at the beginning of the event while only one
hour later the maximum flux at low energy is detected.  On the
other hand, the flux at high energy decreases faster than at low
energy. Only statistical errors are shown.} \label{13eve}
\end{figure}

   In Figure \ref{13eve} is shown the differential
energy spectrum measured with \pam\ in different periods of the
event. It is possible to see that particles were  accelerated
  up to 3-4 GeV.

  A second smaller event occurred  in
conjunction with a X1.5 flare from the same active region (NOAA
10930, $S06^oW46^o$) on Dec 14, superimposing on the Forbush
decrease caused by the Coronal Mass Ejection  of the previous
event reaching Earth. Galactic particle flux thus decreased   in
the energy range up to 3 GeV, whereas solar particles were
accelerated up to 1 GeV for this event. The decrease was also
observed by Wind, Stereo and Polar but not by the GOES satellites,
with the exception of   some variation in the 15-40 MeV channel of
GOES-12 \cite{Mu07}. In case of \pam\ the relative decrease record
by was up to more than $20\%$, extending above 5 GeV.

\begin{figure}[ht]
\begin{center}
\includegraphics[width=1.\textwidth]{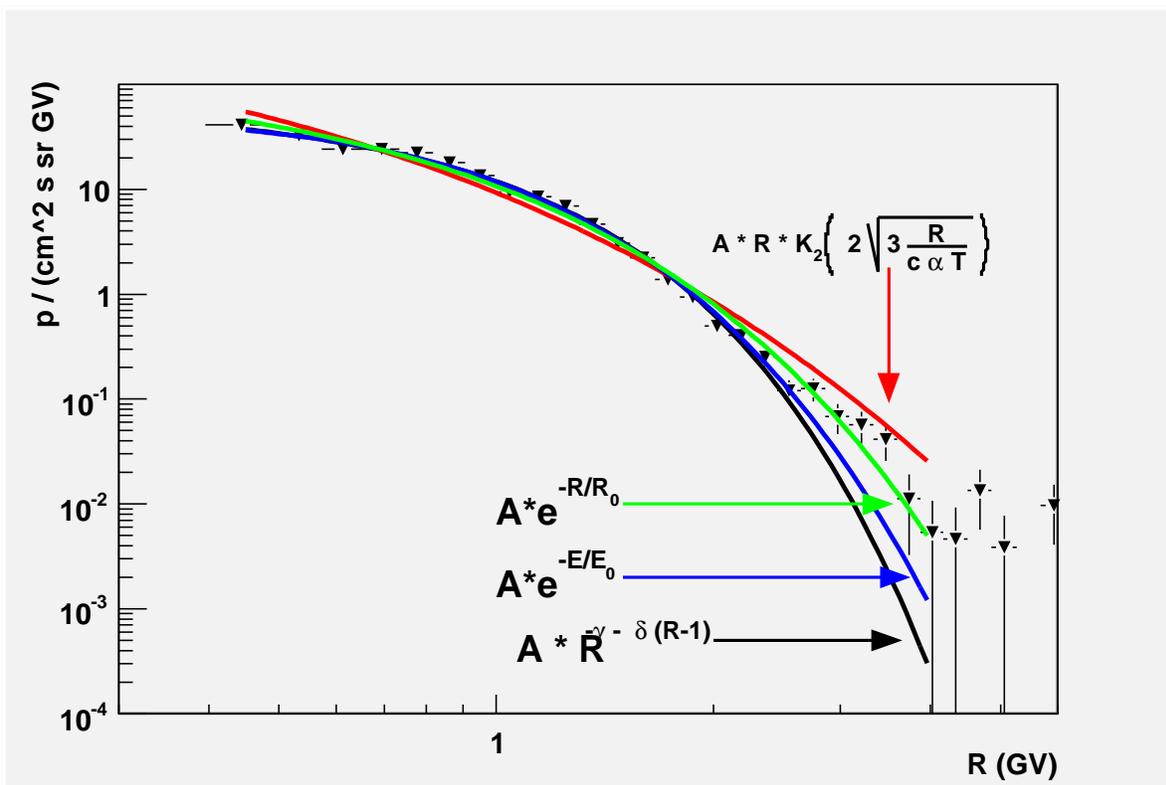}
\end{center}
\caption{Fit of the proton rigidity spectra at the beginning of
the event. Different functions have been employed: from top to
bottom at the high end of rigidity (red: Bessel K2 function, green
Exponential in rigidity,  blue: Exponential in energy, Black
exponential in rigidity. Best fit function according to $\chi^2 $
is the exponential in kinetic energy. ) } \label{confontifits}

\end{figure}

The good energetic resolution and the statistic of the event
allows for a detailed study of the temporal variations of the
spectrum of this event in the energy range 100 MeV - 5 GeV, an
interval  usually not studied in detail with spaceborne  detectors
and usually accessible through ground neutron monitors.   In
Figure \ref{confontifits} is shown the rigidity spectrum  at the
beginning of the event (3:18 - 3:23 GMT 13/12/2006, after
subtraction of the galactic component)
  fitted with various functions (exponential in kinetic energy E, rigidity R, rigidity dependent power law,
    Bessel function):
  \begin{equation}
     \phi = A\: e^{-\frac{E}{E_0}}
     \end{equation}
  \begin{equation}
\phi = A\: e^{-\frac{R}{R_0}}
\end{equation}
  \begin{equation}
\phi = A\: R^{-\gamma - \delta (R-1) }
\end{equation}
  \begin{equation}
  \phi = A\:  p\:  K_2\left( 2\sqrt{ 3 \frac{p}{c \alpha T}
  }\right)
\end{equation}

\begin{figure}[ht]
\begin{center}
\includegraphics[width=1.\textwidth]{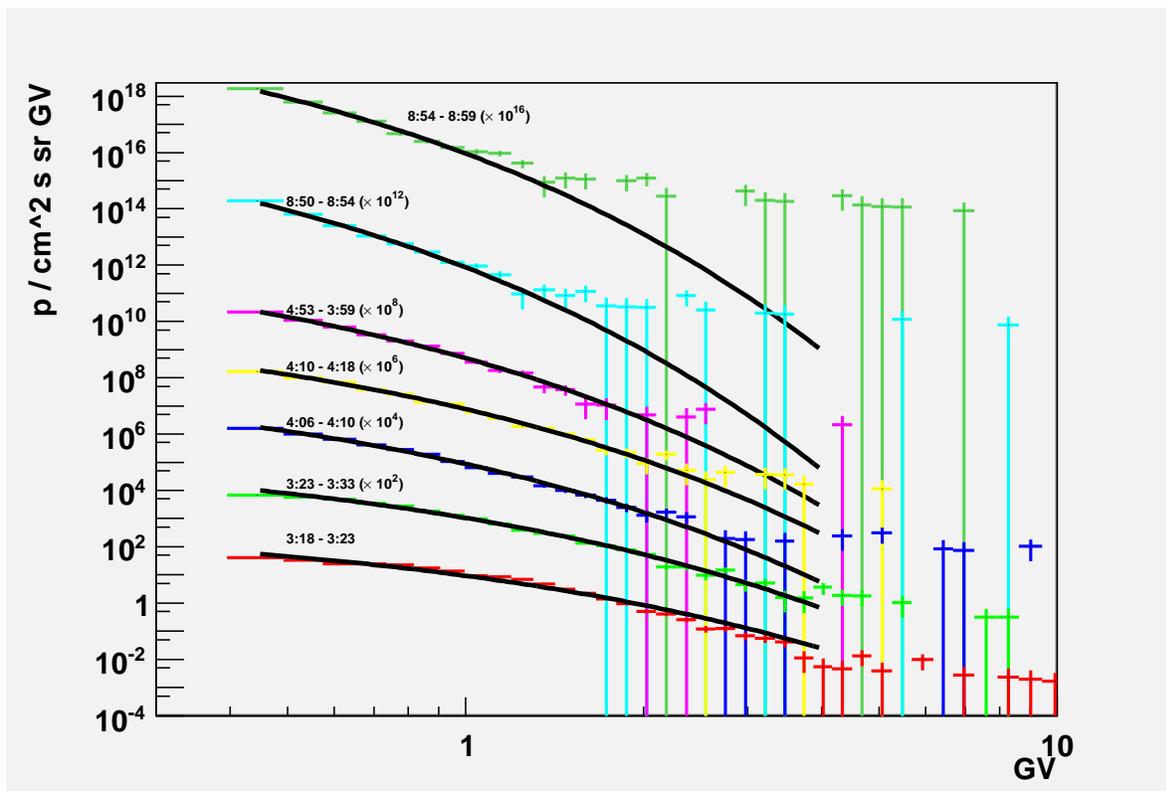}
\end{center}
\caption{Proton differential energy spectra with galactic flux
subtracted in different time intervals during the event of the
13th December 2006.  The fluxes have been scaled in the plot. The
fit with Bessel K2 function. Only statistical errors are shown.}
\label{besselfunc}
\end{figure}

\begin{figure}[!ht]
\begin{center}
\includegraphics[width=1.\textwidth]{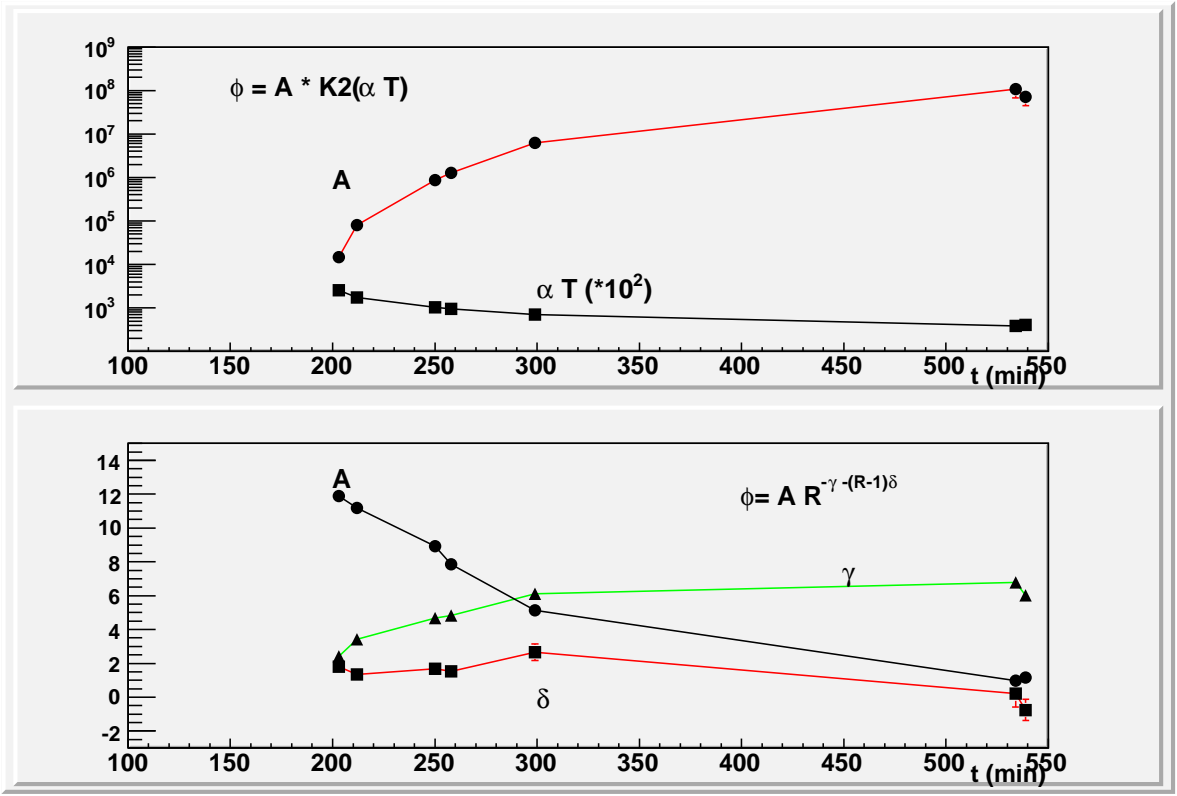}
\end{center}
\caption{Top Panel: Fit of the Rigidity proton spectra with Bessel
K2 function as function of time (minutes from 2006-12-13-00:00
GMT) . Top line: Absolute value. Bottom line: $\alpha T$. It is
possible to see how particle flux increases  and $\alpha T$
decreases with time. Bottom Panel: Fit of the Proton spectra with
Rigidity dependent power law  as function of time ($\phi = A R^{-
(\gamma + \delta (R-1 ))}$) (from top to bottom: $A$, $\gamma$,
$\delta $). In this case particle flux decreases with time. The
decrease of the rigidity dependent term $\delta $ implies a
straightening of the spectrum. } \label{fitpowerlaw}
\end{figure}

The various  functions are representative of   acceleration and
propagation processes under different hypothesis
\cite{mcguirerosevinge1981}.   $K_2 $ is a modified Bessel
function of order 2, with $\alpha T$ as free parameter - $\alpha $
representing an acceleration rate and $T$ the escape time from the
acceleration region, with their product higher in larger events.
This function is a solution of a stochastic Fermi acceleration
\cite{ramaty1979,miller1990,gan1998} in the non-relativistic limit
and therefore is best suited for energies below $\simeq GeV $.
  From the Figure it is possible to see
that all fits are qualitatively good, given the small number of
free parameters, the large energy range of the fit and the
experimental uncertainties. $\chi^2 $ values are high (75, 148,
86, 420 respectively according to the list of formulae above). The
best fit  is obtained for the exponential in energy ($E_0=380
MeV$) which approximates better the shape of the function in the
intermediate energy range. The exponential in rigidity ($R_0=290$
MV) is however better in reproducing the observed spectrum at high
energies. To study the evolution of the particle spectrum of the
event we have fitted the spectra at various times with Bessel
functions (see Figure \ref{besselfunc}) plotting the results of
the fit in Figure \ref{fitpowerlaw}, top panel. With the passage
of time the intensity of the spectrum increases (due to the
increasing flux at low energy) with the lowering  of $\alpha T$
implying a decrease of the acceleration phenomenon. It is
interesting to note that the values obtained are compatible to
what found for other events at lower energies in the energy range
10-100 MeV \cite{mcguirerosevinge1981}, sign that a similar
description of the acceleration and propagation phenomena is valid
at higher energies. Note however that at this stage we do not
infer any conclusion on these processes from the spectral shape of
fit performed. The  same spectra have also been fitted with  a
rigidity dependent power law.   This spectral shape better
describes the decrease of the flux in the lowering of the value of
$A$ with time, shown in Figure  \ref{fitpowerlaw}, bottom panel.
It is also possible to observe   high value of $\delta $ at the
beginning  of the event, implying deviation from a shock
accelerated spectrum. The values of $\delta $ decrease and get
close to zero after some hours   showing how the spectrum becomes
more similar to a single power law and therefore possibly better
representative of a shock accelerated structure. Further studies
will involve other particle species (He, $e^-$) and long term
effects such as the Forbush decrease.

\section{Conclusions}
\pam\ was successfully launched on June 2006. Is  currently
operational in Low Earth Orbit and has recently passed two years
of operation.  We have presented some measurements of protons of
galactic, solar and trapped/secondary origin showing that the
satellite and the detectors are functioning correctly and
providing new data on the particle and antiparticle component of
cosmic rays.

%\begin{thebibliography}{9}
\bibliographystyle{elsart-num}

\begin{table}
\caption{Solar modulation parameter obtained with the fit of the
proton spectrum in different periods. Note the dependence of $\Phi
$ from the assumed  value of the absolute spectrum $A$ . The
variations of $\Phi $ are however more independent from $A $. }
\begin{tabular}{|c|c|c|c|} \hline
   $A$  & 2.0 & 1.95 & 1.91   \\
 p/(cm$^2$ s sr GeV$^{-1}$ GV$^{-2.76}$) & & & \\
 \hline
 &\multicolumn{3}{|c|}{ $\Phi \: (MV)$ } \\
\hline  July 2006  & 635 $\pm $ 3 & 627 $\pm $ 2 & 621 $\pm $ 3 \\
 August 2007 & 565 $\pm $ 2 & 558 $\pm $ 3 & 552 $\pm $ 3 \\
February 2008 & 561 $\pm $ 3 & 553 $\pm $ 2 & 546 $\pm $ 2  \\
\hline
\end{tabular}
 \label{solmod}
\end{table}

\begin{table} \small
\caption{Fit of the core regions of the SAA according to a
rigidity dependent power law spectrum. } \label{fitsaa}
\begin{tabular}{|c|c|c|c|c|}
\hline Region & A & $\gamma $ & $\delta
$ & $\chi^2$  \\
 $G$ & $p/cm^2 s\:  sr GeV$ & &  $GV^{-1}$ & $/ndf$ \\
\hline
$0.19 < B$               & 4.0 $\pm $ 0.4 & 3.6 $\pm $ 0.6  & 2.0 $\pm $ 0.4 & 0.04 \\
$  0.19 < B < 0.20$    & 1.0 $\pm $ 0.1 &  4.6 $\pm $ 0.8 & 1.6 $\pm $ 0.5 & 0.17 \\
$ 0.20 < B < 0.21 $   & 0.05  $\pm $ 0.008  & 5.6 $\pm $
1.5 & 0.8 $\pm $ 1.4 & 0.40 \\
\hline
\end{tabular}
\normalsize
\end{table}

\end{document}